\documentclass{aastex}          
\usepackage{spr-astr-addons}    
\usepackage[]{natbib}
\usepackage{color}
\usepackage{epstopdf}
\usepackage{dblfloatfix}
\usepackage{amsmath}
\begin{document}
%
\title{Near Infrared High Resolution Spectroscopy and Spectro-astrometry of Gas in Disks around Herbig Ae/Be Stars}

\shorttitle{<Short article title>}
\shortauthors{Brittain et al. 2014}

\author{Sean D. Brittain \altaffilmark{1}} 
\author{Joan R. Najita \altaffilmark{2}} 
\and 
\author{John S. Carr \altaffilmark{3}}

\email{sbritt@clemson.edu} 

\altaffiltext{1}{Clemson University}
\altaffiltext{2}{National Optical Astronomy Observatory}
\altaffiltext{3}{Naval Research Labs}

\begin{abstract}
In this review, we describe how high resolution near  infrared spectroscopy and 
spectro-astrometry have been used to study the disks around Herbig~Ae/Be 
stars. We show how these tools can be used to identify signposts 
of planet formation and elucidate the mechanism by which Herbig Ae/Be 
stars accrete. We also highlight some of the artifacts that can complicate 
the interpretation of spectro-astrometric measurements and discuss best 
practices for mitigating these effects. We conclude with a brief discussion 
of the value of long term monitoring of these systems. 
\end{abstract}


\section{Introduction}
\label{sec:intro}
Herbig Ae/Be stars are intermediate mass (2-8M$_{\sun}$) pre-main 
sequence and zero age-main sequence stars whose spectral energy 
distribution reveals an infrared excess (Herbig 1960; Waters \& Waelkens 
1998; Vieira et al. 2003). They are the precursors to main sequence 
A-stars many of which have debris disks  (Su et al. 2005; 2006). As 
the higher mass analogs to classical T Tauri stars, Herbig Ae/Be stars 
provide an important link between their lower mass solar analogs and high 
mass stars. 

A great deal of progress has been made in the study of the circumstellar 
environment of Herbig Ae/Be stars since the last international conference 
on Herbig Ae/Be stars (The et al. 1994a). Over the past two decades, 
advances in coronagraphic imaging (Grady et al. 1999, 2005, 2014; 
Weinberger, et al. 1999), 
the (sub)millimeter (Mannings \& Sargent 1997; Dent et al. 2005) and infrared observations 
(e.g. Meeus et al. 2001, 2012), and SED modeling (Dullemond et al. 2007) have 
advanced our understanding of this environment. While the the geometry 
of the dust giving rise to the infrared excess was controversial two decades 
ago (The et al. 1994b), it is now clear that Herbig Ae/Be stars possess flared 
disks in a Keplerian orbit - usually with a puffed up inner rim. Additionally, 
the study of post-main sequence stars that have evolved from A stars indicate that these disks form giant planets 
perhaps even more prodigiously than their lower mass counterparts (Johnson 
et al. 2010). 

In addition to these tools, near infrared spectroscopy provides valuable 
insight to the gas in the inner region of the disk. For example, while this 
region ($\lesssim$50~AU) remains difficult to study by direct imaging, the 
kinematic profile of resolved emission lines can serve as a surrogate for 
high resolution imaging by providing spatial information about the gas 
on milliarcsecond scales assuming the gas is in a Keplerian orbit and that 
the stellar mass and disk inclination are known. Ground based high resolution 
(R$\gtrsim$20,000) spectroscopy of ro-vibrational CO lines have been 
used to measure the distribution of molecular gas relative to dust in the 
inner disk. This technique was pioneered in the study of classical T Tauri 
stars and embedded YSOs (e.g. Najita et al. 2000, 2007) and has been 
applied extensively to gas in disks around Herbig Ae/Be Stars (e.g. 
Brittain \& Rettig 2002; Brittain et al. 2003, 2007; Blake \& Boogert 2004; 
Goto et al. 2006;  van der Plas et al. 2009, 2010; Salyk et al. 2009; 
Brown, J. et al. 2013).  The study of ro-vibrational CO lines has also complemented 
FIR studies of disks around Herbig Ae/Be stars by adding parameter space to 
constrain thermochemical models of disks (e.g. Carmona et al. 2014; Thi et al. 2014).

Spectroscopy of the infrared \ion{H}{1} recombination lines has also 
advanced our understanding of Herbig Ae/Be stars. Moderate resolution 
spectroscopy combined with interferometry has constrained the size of 
the emitting region of \ion{H}{1} to typically within $\sim$0.1~AU (Eisner 
et al. 2009; Kraus et al. 2008). A great deal of work has also been done 
to calibrate these lines against the accretion luminosity so that they can 
be used to measure the accretion rate of Herbig Ae/Bes (van den Ancker 
et al. 2005; Garcia Lopez et al. 2006, Donehew \& Brittain 2011; Mendigut{\'{\i}}a 
et al. 2011; Pogodin et al. 2012).

A particularly exciting development in the use of NIR spectroscopy to 
study gas in disks is the application of spectro-astrometry. Spectro-astrometry is 
the measurement of the spatial center of the point spread function (PSF) of 
a spectrum as a function of wavelength (Bailey et al. 1998a).
Because variations in the center of the PSF can be measured on much 
smaller angular scales than the diffraction limit of an instrument
spectro-astrometry can provide spatial 
information on sub-milliarcsecond scales. This technique was initially 
developed with the advent of CCDs in the 1980s (see Bailey 1998b for 
a review) and has since been applied extensively to extract additional 
spatial information from spectra of a number of astrophysical phenomena 
such as binaries (Bailey et al. 1998a; Porter et al. 2004), YSO jets (Takami 
et al. 2003, Whelan et al. 2004; Davies et al. 2010), and circumstellar disks 
(Acke \& van den Ancker 2006; Pontoppidan et al. 2008; Brittain et al. 2013; 
Brown, L. et al. 2013). 

 In this review we discuss how modeling line profiles spectrally and 
 spatially provide information about the distribution of gas in disks 
 around Herbig Ae/Be stars. We then highlight two example applications: 
 identification of signposts of planet formation and the study of the origin 
 of \ion{H}{1} emission lines in Herbig Ae/Be stars. We follow with a 
 discussion of sources of potential artifacts that may complicate the 
 interpretation of spectro-astrometry measurements. We show the scope of these artifacts 
 and how to acquire the data so as to mitigate their effect. We conclude 
 with a brief discussion of future prospects for ground based study of gas in disks. 

\section{Technique}
\subsection{Line Profile Modeling}
Planet forming disks begin 
their lives 
with high optical depths at most wavelengths. For the minimum 
mass solar nebula, the surface density at 1~AU is 1700~g~cm$^{-2}$ or 
about N(H)=10$^{27}$~cm$^{-2}$. For a gas to dust ratio and grain size 
distribution representative of the local interstellar medium, only the top 
10$^{-5}$ of the disk is optically thin at 5$\mu$m. Grain growth and settling 
increase the depth of the optically thin portion of the disk atmosphere, 
but even so NIR observations only probe a thin layer of the disk. The line 
emission observed in 
the 
disk arises from this thin warm layer of the atmosphere. 
Thus observing H$_2$ can be very challenging because of the weak 
oscillator strengths of the quadrupole transitions (Bitner et al. 2007; 
Carmona et al. 2011)
and large H$_2$ column densities are needed for detectable emission. 
CO has proven to be the most commonly observed 
species in disks despite the fact that its abundance is 10$^{-4}$ -- 10$^{-6}$ 
that of H$_2$. The ro-vibrational transitions of CO are a useful tracer of gas 
for a number of reasons. First, the relatively low sublimation temperature and 
high dissociation temperature makes it abundant throughout a circumstellar 
disk. Second, even relatively small columns of CO are self shielding. 
Ro-vibrational transitions are also readily observable from the ground 
(moderate Doppler shifts are necessary for the J$\lesssim$11 v=1--0 transitions). 

We can trace the distribution of this warm layer of gas in the 
disk by modeling the kinematic profile of the lines. If we assume 
the gas follows a circular Keplerian orbit and know the disk inclination 
and stellar mass, then the line profile can be readily calculated (e.g. 
Smak 1981; Figure 1). If the line is resolved, the inner and outer edge 
of the gas emission can be calculated. The half width at zero intensity 
(HWZI) of the emission line corresponds to the projected velocity of 
the gas at the inner edge of the detected region. The half-width of the peak 
separation corresponds to the velocity of outer extent of the emitting region.


 \begin{figure}[h]
 \includegraphics[width=\columnwidth]{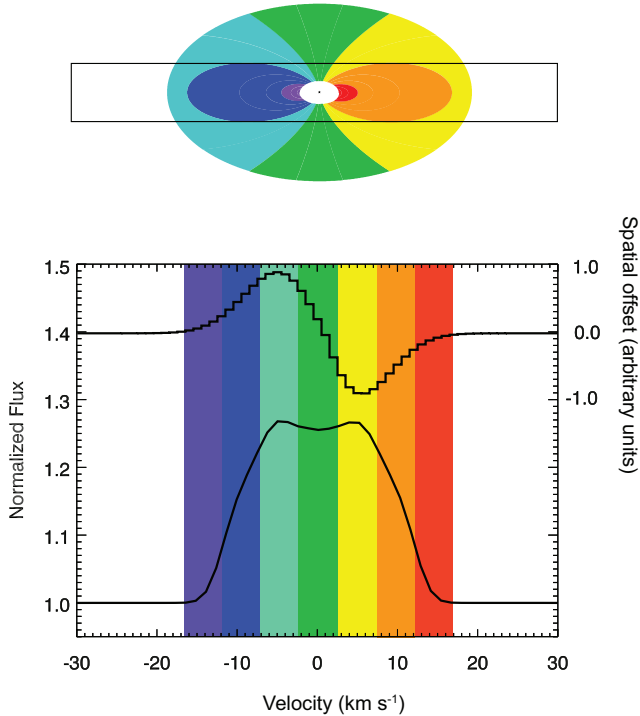}
 \caption{Schematic of a rotating disk, emergent spectral line and spectro-
 astrometric signal. In the top panel the projected isovelocity contours are 
 colored. The corresponding parts of the emission line are given the same 
 color. The half width at zero flux corresponds to the velocity of the gas at the 
 inner edge of the disk. The half-width of the peak separation corresponds to 
 the velocity of outer extent of the emitting region. When the slit of the 
 spectrograph is aligned with the semi-major axis of the disk (the rectangle 
 superimposed on the schematic), the blue shifted gas is spatially offset 
 toward one end of the disk and the red-shifted gas is spatially offset toward 
 the other end of the slit. The magnitude of the offset depends on the spatial
  extent of the gas, the radial distribution of the disk flux, and the relative 
  brightness of the continuum. } 
 \label{fig:1}
 \end{figure}

Generally the data are modeled by assuming the gas distribution is axisymmetric and follows circular 
Keplerian orbits, however, this may not accurately describe gas in circumstellar 
disks. For example, the gas in orbit may be elliptical giving rise to asymmetric 
line profiles (Reg{\'a}ly et al. 2010) or the gas may arise in a non-Keplerian orbit such 
as a disk wind (Carr 2007; Pontoppidan et al. 2011). 

\subsection{Spectroastrometry}
\label{sec:Observations}
Using a relatively new technique, 
spectroastrometry, one can test the assumption that the gas arises in 
a Keplerian orbit.  
Spectro-astrometry has been described in a number of papers (e.g. Bailey 1998ab; 
Porter et al. 2004; Brannigan et al. 2006; Whelan \& Garcia 2008), 
so we limit our discussion to a brief overview of this technique - particularly as it 
applies to circumstellar disks.

While the spatial resolution of a PSF is limited to $\sim$1.2D/$\lambda$, 
the centroid of the PSF can be measured to a small fraction of that. For a well sampled Gaussian
PSF dominated by photon noise, the center of the Gaussian can be measured to an accuracy of, 

\begin{equation}
\delta b \sim 0.4 \frac{FWHM}{SNR}
\end{equation}

\noindent where $b$ is the center of the Gaussian, $FWHM$ 
is the full width at half maximum power of the Gaussian, and $SNR$ is the 
signal to noise ratio of the PSF (Brannigan et al. 2006). With a SNR of 100 and an AO corrected beam of 0$\arcsec$.12, 
the centroid of the PSF can be measured to an accuracy of 0.5 milliarcseconds
assuming the PSF is adequately sampled. The spectro-astrometric
signal is not a direct measure of the extent of the gas, rather it is the projection of the center of 
light of a given velocity channel along the axis of the slit. 

Typically, the 
continuum is much brighter 
than the emission lines, so the continuum dominates the center of light.  
The extent of the offset depends on the angular extent 
of the emission, the asymmetry of the distribution of the emission 
along the axis of the slit, the brightness of the emission relative to the continuum,
and the relative position 
angle of the disk and the slit. For example, if the slit is aligned with the semi-major axis of an 
inclined axisymmetric disk, gas in a Keplerian orbit will show emission lines where the red shifted
side of the line is spatially offset toward one end of the slit while the blue shifted side of the emission line
is offset toward the opposite end of the slit (figure 1). If the slit is rotated so that it is aligned with the semi-minor axis
of the disk then the center of light of the red and blue shifted gas will be aligned with the center of the PSF and 
no offset will be measured.

This technique has provided 
valuable information about gas in disks.  For example, high resolution spectra of the
 [\ion{O}{1}]$\lambda$~6300~\AA\  lines have been 
 analyzed spectro-astrometrically (Acke \& van den Ancker 2006). These authors
 show that the line is consistent with gas in Keplerian orbit 
 about Herbig Ae/Be stars rather than arising in an outflow as
 in the case of T Tauri stars. In contrast, spectro-astrometric 
 analysis of high resolution spectra of ro-vibrational CO emission
 from strongly accreting YSOs shows that the gas is less extended 
 than the narrow profiles indicate if the gas were in a Keplerian 
 orbit (Pontoppidan et al. 2011; Bast et al. 2012). 

In the section that follows we describe two applications of spectral 
profile fitting and SA to elucidate aspects of disks around Herbig Ae/Be stars.

\section{Applications}
\subsection{Signposts of Planet Formation}
The identification of gas giant planets forming in 
circumstellar disks will represent a major 
milestone in the study of planet formation. 
By identifying forming planets we can connect the initial conditions 
in disks to the kinds of planets that form. We can also determine 
the disk radii at which planets form for comparison with 
the orbital radii of planets observed in more mature systems. 
Such observations can also test theoretical predictions that 
planets clear gaps in disks and accrete mass through circumplanetary 
disks, processes that affect the evolution of planetary masses.
This approach is complementary to the direct detection of 
forming gas giant planets. 

Direct detection is challenging due to the 
small angular separation between the forming gas giant planet and 
the host star as well as the high 
contrast between the forming planet and the disk. 
However, several exciting candidate forming objects have been identified 
by direct imaging (e.g., T~Cha Hu\'{e}lamo et al. 2011; LkCa~15 Kraus \& Ireland 2012; 
HD~100546 Quanz et al. 2013). 

 \begin{figure}[h]
 \includegraphics[width=\columnwidth]{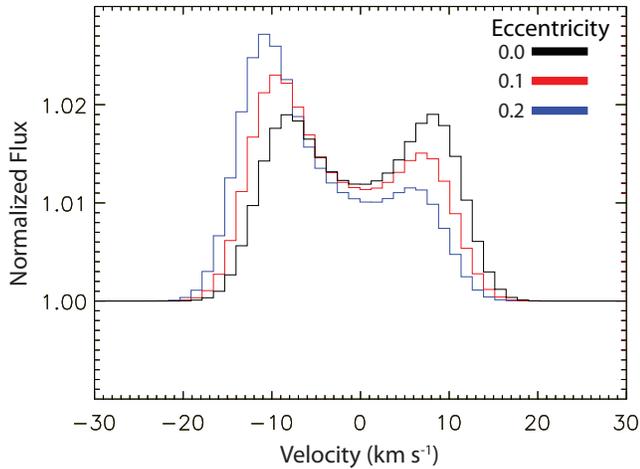}
 \caption{Synthetic line profile from gas arising from an eccentric annulus. The 
 line profile of a line arising from an annulus is plotted. The asymmetry of the 
 line changes with the eccentricity of the disk. The asymmetry arises from two 
 effects. First, the gas near periastron is moving faster than gas near apastron. 
 Second, the gas at periastron is closer to the star and thus brighter than the more 
 distant gas. Even if the line is not resolved, this effect can be inferred from the 
 shift of the centroid of the line relative to the systemic velocity of the system. } 
 \label{fig:2}
 \end{figure}

Another approach to identifying forming gas giant 
planets is to identify dynamical markers of tidal interactions between the disk and 
forming object. Hydrodynamic modeling of planet disk interaction indicates that a 
planet with a mass greater than $\sim$3M$_{\rm Jupiter}$ on a circular orbit can induce an 
eccentricity in the disk in the vicinity of its orbit (e.g. Kley \& Dirksen 2006). 
Simulations show that the eccentricity can be as high as 0.25 and falls off 
as approximately r$^{-2}$. The profile of a line arising from an eccentric annulus will generally not be 
symmetric. The gas at periastron will be faster than the gas at apastron and 
generally the temperatures will differ as well. This results in a distinctive asymmetric 
line profile (figure 2; see Reg{\'a}ly et al. 2010 and Liskowsky et al. 2012). Interestingly, 
the semi-major axis of the disk precesses very slowly ($\sim$10${\degr}$ 
per 1000 orbits), so the line profile 
should remain constant over decadal timescales. Such a line profile 
has been observed in OH spectra from the Herbig Ae binary V380~Ori (Fedele et al. 2011) 
and the Herbig Be star HD~100546 (Liskowsky et al. 2012). 
In the case of V380~Ori, it is likely a stellar companion that is driving the 
eccentricity whereas in the case of HD~100546 it is likely that a 
substellar object is driving the eccentricity. In principle, an asymmetric line profile could
also arise from the uneven sampling of a resolved axisymmetric disk 
(e.g. Hein Bertelsen et al. 2014) which we will discuss in section 3.

 \begin{figure*}[!t]
\begin{center}
 \includegraphics[width=6.5in]{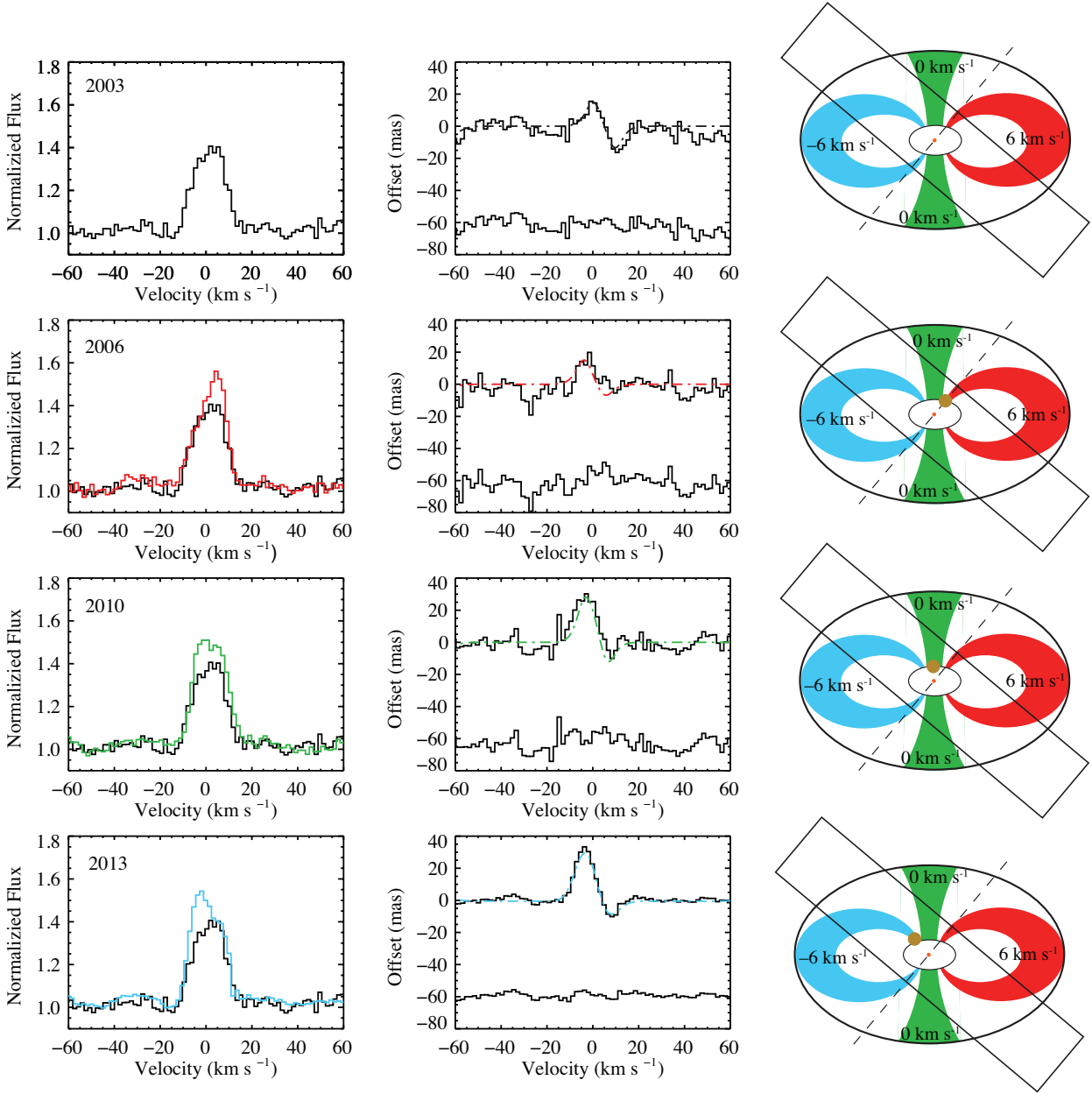}
\end{center}
 \caption{Evolution of CO emission line profile and spectro-astrometric signal (Brittain et al. 2014). 
 In the first column the v=1--0 P26 CO line observed in 2003 is plotted (black) 
 and compared to the same line observed in 2006 (red), 2010 (green), and 
 2013 (cyan). The Doppler shift of the excess emission can be seen in this column.
 In the second column, the spectro-astrometric signal is plotted 
 for each epoch along with the model (dot-dashed line) and the residual. 
 The spectro-astrometric signal is modeled by adding an 
 extra source of emission in the inner edge of the disk at the position 
 corresponding to the Doppler shift of the excess (illustrated in the third 
 column). While the inner edge of the disk is roughly 130$mas$ from the 
 star, the center of light of the CO emission is offset from the center of the 
 PSF $\pm$15$mas$ along the axis of the slit. The extra source of emission 
 in orbit around the star causes the center of light to shift.  This source of emission
 was hidden by the circumstellar disk in 2003.  } 
 \label{fig:3}
 \end{figure*}

Another way to identify ongoing planet formation is by detecting 
emission arising from a circumplanetary disk feeding the planet. 
Circumplanetary disks can be discovered 
by their line emission (Brittain et al.\ 2013; 2014) 
and also potentially by their dust continuum emission 
(Zhu 2014; Isella et al.\  2014).  
Simulations of gas giant formation 
indicate that the disk feeding the gas 
giant planet will have a radius roughly one-third of the Hill Sphere 
(Quillen \& Trilling 1998; Ayliffe \& Bate 2009ab; Martin \& Lubow 2011), 
though the full accretion envelope is likely intrinsically three dimensional with complicated flow patterns 
(e.g. Tanigawa et al. 2012; Ayliffe \& Bate 2012; Gressel et al. 2013).   Thus a 5M$\rm_{Jupiter}$ 
planet forming at an orbital radius of 10~AU should have a 
circumplanetary disk with a radius of 0.3~AU - comparable to the size of the
emitting region of ro-vibrational CO emission observed toward classical T Tauri stars (Najita et al. 2003). 
 
In the case of Herbig Ae/Be 
stars with transition disks, hot bands of ro-vibrational 
CO emission are observed from R$\gtrsim$10~AU (e.g. HD~179213, van der Plas 2010; 
HD~100546, Brittain et al. 2009; HD~141569 Brittain et al. 
2003; HD~97048, van der Plas et al. 2009; and Oph~IRS~48, 
Brown et al. 2012) due to the UV fluorescence of the gas. To detect the 
IR ro-vibrational emission lines that result from this process, a large 
area is needed because these transitions are exceeding optically thin.
For example, the line flux of CO emission arising from HD~100546 is 
comparable to that observed from classical T Tauri stars in Taurus 
(compare for example, Brittain et al. 2009 and Najita et al. 2003). 
However, the emitting region of the CO observed toward HD~100546 
extends over 50~AU while the emission from the classical T Tauri 
stars typically originates within 1~AU of the star. Thus UV fluoresced gas
originating from a disk with a radius less than 1~AU will reveal 
ro-vibrational CO lines with luminosities 3-4 orders of magnitude fainter 
than what is observed from the transition disks mentioned above. 

The contrast between the gas conditions
which produce UV fluorescence (broadly distributed, perhaps tenuous, gas) and collisionally 
excited emission (warm, dense material) provides a way to identify 
circumplanetary disks.  The surface density and temperature of the circumplanetary disk is expected to be 
enhanced relative to the feeding material (Klahr \& Kley 2006; Matin \& Lubow 2011). If the gas 
is of order $10^3$~K and optically thick, then the flux of the emission from the circumplanetary disk will be 
similar to what is observed from classical T Tauri stars. 

A Herbig Ae/Be star with a transition disk created by a 
forming gas giant planet should then give rise to
v=1--0 lines that are a blend of emission from the circumstellar disk and 
circumplanetary disk while the hotband transitions will be overwhelmingly 
dominated by the UV fluoresced gas in the circumstellar disk. Thus as 
the circumplanetary disk orbits the star, the shape of the v=1--0 
lines will change with the motion of the circumplanetary disk. The hotband 
lines provide a constant fiducial profile against which to 
measure the evolution of the v=1--0 lines. This behavior, variable  v=1--0 
profiles with constant hotband profiles, is observed in 
HD~100546 (Brittain et al. 2013; 2014). 

In January 2006 the red side of the v=1--0 P26 line at 0--10~km~s$^{-1}$ 
brightened relative to what was observed in January 2003 (second row of figure 3). The Doppler shift
of this excess emission was 6$\pm$1~km~s$^{-1}$. In December 2010, the v=1--0 P26 was still 
brighter than it was in January 2003, however, the Doppler shift was --1$\pm$1~km~s$^{-1}$ (third row of figure 3).
In March 2013 the excess emission was Doppler shifted --6$\pm$1~km~s$^{-1}$. Assuming the source
of the excess emission is in a Keplerian orbit, the projected Doppler shifts and epochs of the observations
place the emission near the inner edge of the outer disk (R$\sim$13~AU). One way 
to test whether the shift of the excess v=1--0 line emission is due to gas in a 
Keplerian orbit and determine the radial position of the emission is to use spectro-astrometry.

Consider an inclined axisymmetric disk of gas 
in a circular Keplerian orbit. If the slit is aligned
near the semi-major axis of the inclined disk, then the blue side of the 
emission lines formed in the disk will be offset in one direction along 
the slit axis and vice versa for the red side of the line, and the offsets 
would be symmetric (figure 1). This was observed for the v=1--0 P26 line in 2003 
(top row of figure 3). But what happens 
if we add an extra localized source of emission at a given velocity 
to our disk? In this case, the center of light at the velocity of the extra source
 will be shifted closer to the added source of emission. 

The extent and direction of the 
shift will depend on the location of the center of light along the slit axis. 
In the case of HD~100546, the first three observations were taken with 
the slit in its default position angle of 90$\degr$ (Brittain et al. 2013). 
The position angle of the disk is 140$\degr$ (Ardila et al. 2007). In 2006 
not only was the red side of the line brighter 
relative to its brightness in 2003, the center of the PSF of the red side 
of the line was less extended in 2006 than it was in 2003 (second row of figure 3). 
As it turns out, for the combination of the slit PA and disk PA,
an excess source of emission that orbits close to the inner disk edge of the outer disk (R$\sim$13~AU)
with a projected velocity of 6~km~s$^{-1}$ will be located near 
the center of the slit, just as required to explain the spectro-astrometric signal 
(second row of figure 3). In 2010, the 
spectral profile of the extended emission
was broader and shifted to $-1\pm$1~km~s$^{-1}$. Placing this emission 
in the disk near the inner rim at the orbital position corresponding to this shift
reproduces the spectro-astrometric signal (third row of figure 3). The same is 
observed for the observations acquired in 2013. From 2006 to 2010 to 2013, the Doppler 
shift of the emission, location of the emission projected along the slit axis, 
and period of the emission are all consistent with a source of thermal CO 
emission in a circular Keplerian orbit near 13~AU (Brittain et al. 2014). 

\subsection{The Origin of \ion{H}{1} lines}
A second important application of spectro-astrometry to the study of Herbig 
Ae/Be stars is the study of the origin of the \ion{H}{1} recombination lines. 
The origin of these lines remains uncertain (Kraus et al. 2008): 
do they form in the stellar wind, disk wind, decretion disk, accretion flow, 
or some combination thereof? If the \ion{H}{1} lines are connected 
to accretion, then they may
 serve as a useful proxy of the stellar accretion rate (e.g. Garcia Lopez et al. 
 2006). Several studies have 
 attempted to calibrate various emission line diagnostics to 
the accretion luminosity (e.g. Donehew \& Brittain 2011; Mendigut{\'{\i}}a et al. 
2011, 2012, 2013; Oudmaijer et al. 2011; Pogodin et al. 2012;  see also 
Salyk et al. 2013). Typically, these studies convert the veiling of 
the Balmer discontinuity to an accretion luminosity using a magnetospheric 
accretion model (e.g.  
Muzerolle et al. 2004) and compare this to the luminosity of various emission 
lines.  
For example, Donehew \& Brittain (2011) apply this method and 
find that the relationship between the luminosity of Br$\gamma$ and the 
accretion luminosity determined for Herbig Ae stars is consistent with the 
relationship found for classical T Tauri 
stars (Muzerolle et al. 1998) and intermediate mass T Tauri stars (Calvet et 
al. 2004); however, the Herbig Be stars 
do not follow the same trend. 

 \begin{figure*}[!ht]
 \includegraphics[width=6.5in]{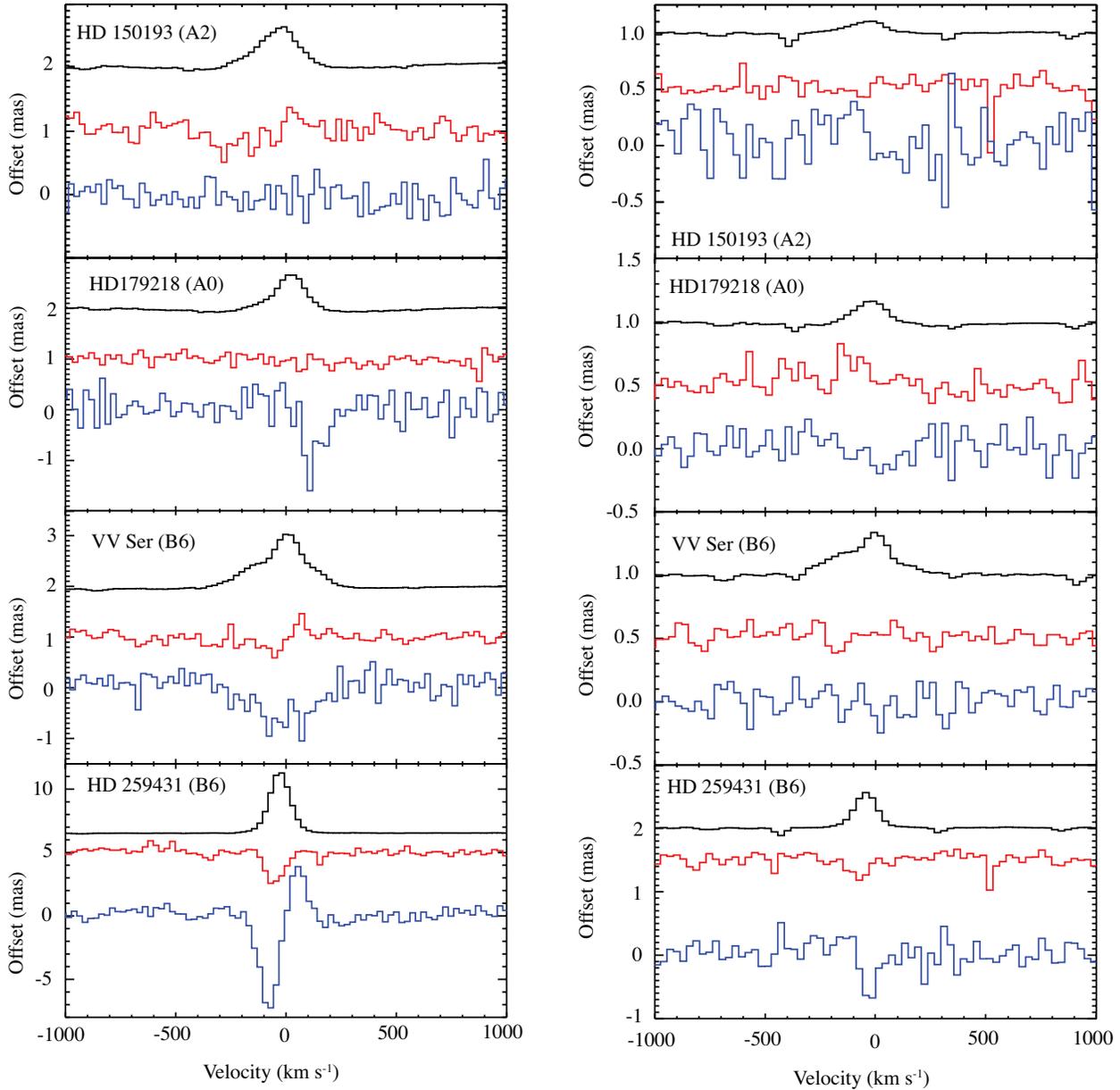}
 \caption{Spectro-astrometric measurement of Pa$\beta$ (left column) and 
 Br$\gamma$ (right column) with NIFS on Gemini North. The spectral profile
 of the line is plotted (black) above the spectro-astrometric signals. The
 red lines correspond to the centroid measurements made along the slitlet (the x-direction) and the 
 blue lines refer to the centroid measurements made perpendicular to the slitlet (the y-direction). No significant
 signal is detected for the HAe stars, but a signal is detected for the HBe stars. 
 Modeling the line profiles and signals is an ongoing project (Adams et al. 2015 - in prep.).} 
 \label{fig:4}
 \end{figure*}

Whether the correlation between the luminosity of 
the 
\ion{H}{1} emission lines 
and 
the inferred accretion luminosity is physical or fortuitous 
remains to be seen. 
One way to 
investigate this issue 
is to measure the variability of the veiling 
of the Balmer discontinuity and emission line luminosities
simultaneously (Mendigut{\'{\i}}a et al. 2013). With the advent of broad coverage 
spectrographs  
such as X-Shooter on the VLT (covering $\sim0.3-2.4\micron$), it is now 
possible to simultaneously observe the 
veiling of the Balmer discontinuity and the emission line 
diagnostics that fall within this range (Oudmaijer et al. 2011; Mendigutia et al. 2014). 
Ongoing studies along these lines will be 
able to determine whether or not the veiling and line emission are
connected.   However, they will not be able to establish that Herbig Ae/Be 
stars accrete magnetospherically - as assumed in the calculation of 
the accretion luminosity - or identify where the  \ion{H}{1} emission lines originate. 

At first blush the assumption that Herbig Ae stars accrete 
magnetospherically would seem unjustified. Herbig Ae/Be 
stars are not fully convective thus one would not expect 
that they would generate the strong kilogauss magnetic fields 
characteristic of their lower mass analogs (Johns-Krull 2007). 
Indeed, studies of the magnetic properties of Herbig~Ae 
stars indicate that they do not possess strong well ordered 
surface fields (Alecian et al. 2014). On the other 
hand red shifted hydrogen lines are observed in Herbig~Ae 
stars with velocities of several hundred km s$^{-1}$ 
indicative of material in free fall onto the star (the tell tale 
signature of magnetospheric accretion; 
Guimar{\~a}es et al. 2006).  Thus the assumption that at least 
some Herbig Ae/Be stars accrete magnetospherically is 
not without warrant. 

\begin{figure*}[!b]
 \includegraphics[width=6.5in]{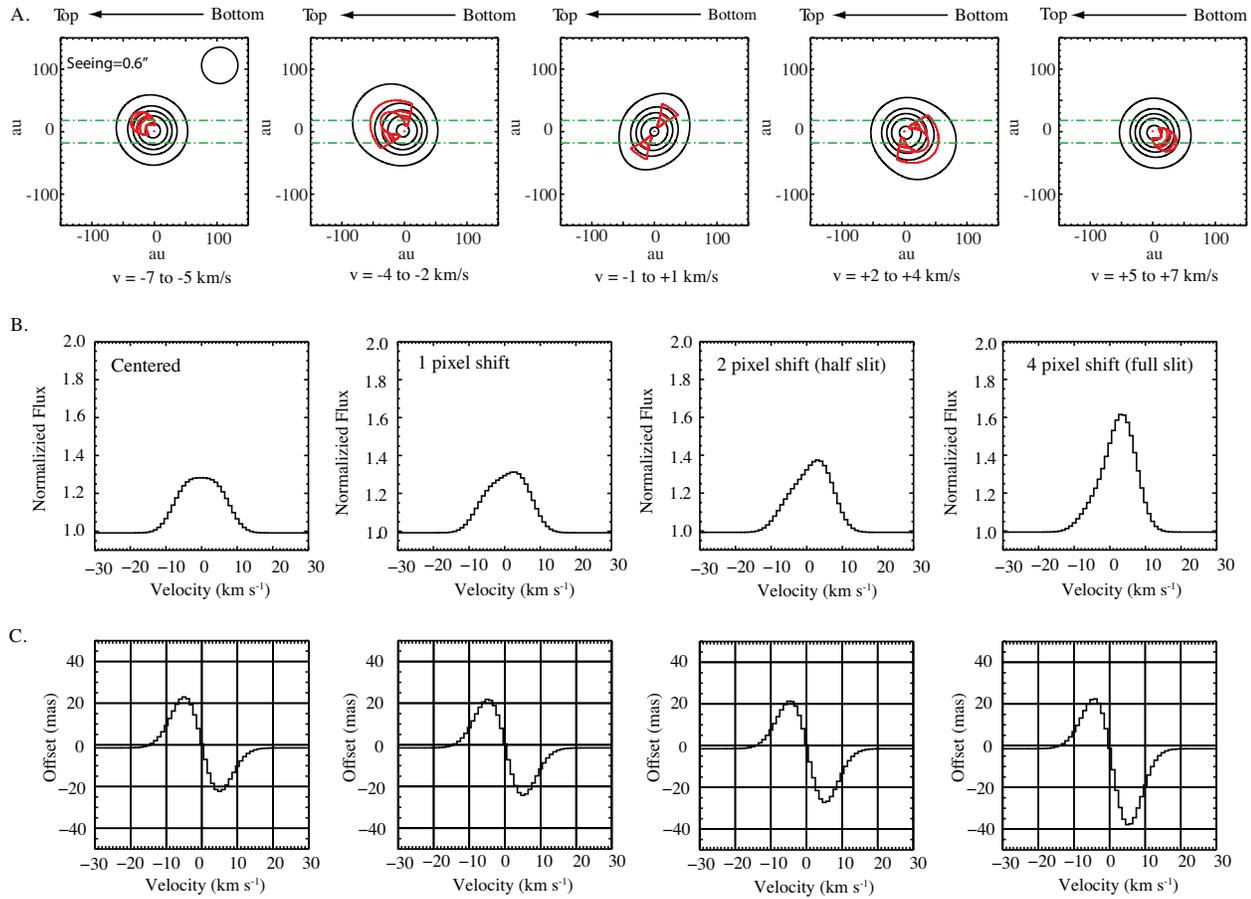}
 \caption{Illustration of artifacts resulting from pointing errors. In row A, the
 isovelocity contours are plotted in red. Also plotted are the same contours
 convolved with two dimensional Gaussian profile with a full width at half max
  equal to 0$\arcsec$.6. The portion of the disk sampled by the slit on the
  Phoenix spectrograph is marked with green dot-dashed lines. In row B, the 
  resultant line profile arising from the disk is plotted. In the first panel, the PSF 
  of the disk is centered in the slit. In the next the center of the PSF is offset by 
  one pixel, and so on. In row C, the spectro-astrometric measurement of the 
  PSF is plotted. Each panel corresponds to the offset noted in the row above. 
  When the disk emission is resolved and the PSF is not properly centered in the
  slit, the profile is asymmetric. This has a minor effect on the line profile and 
  spectro-astrometric signal unless the offset is substantial. Misaligning the center 
  of the PSF and slit results in about a 20\% error when offset is half a slit width 
  and 50\% error when the PSF is offset a full slit width. With care the center of the 
  PSF can be measured and placed in the center of the slit with an accuracy of about 
  0.5 pix. } 
 \label{fig:5}
 \end{figure*}

Before the recombination \ion{H}{1} lines can be used reliably as a proxy for 
the stellar accretion rate, it is crucial that we understand the 
physical origin of these lines. One promising development is the use of 
spectro-astrometric 
measurements of \ion{H}{1} emission using integral field spectrographs. This 
method has recently been applied to massive YSOs.  Davies et al. (2010) 
measured 
Br$\gamma$ emission from the massive YSO W33A with NIFS on Gemini 
North. They 
were able to measure the centroid of the PSF with an accuracy of 0.1 ~$milli
$arcseconds and show that the line originates in an outflow.

Application of the same method to Herbig Ae/Be stars is beginning to show 
promising results. For example 
(Adams et al. 2015), show they can achieve a root 
mean square fidelity of 100$\mu$arcseconds 
with their observation of Herbig Ae/Be stars. This is within a factor of a few 
of being able to distinguish emission from a magnetosphere, outflow, and 
gas in a Keplerian 
orbit in the inner 0.1AU. A study of four Herbig Ae/Be stars finds SA signatures 
from \ion{H}{1} 
from the HBe stars but not the HAe stars in the sample (Figure 4). 
By combining these data we are
also able to show that our fidelity is limited by photon noise. Increased 
integration time should 
provide a means to improve the sensitivity of these measurements by a factor 
of a few and 
thus discriminate between a wind, disk, or funnel flow origin for the nearest 
Herbig Ae/Be stars.  
Making such high precision measurements requires attention 
to potential artifacts in the data.

\section{Pitfalls}
Several authors have summarized best practices 
for using SA to study 
phenomena such as binaries and outflows 
(Bailey et al. 1998; Baines et al. 2004; Porter et al. 2004; Brannigan et al. 
2006; Whelan \& Garcia 2008). Because SA involves making high 
precision measurements of the center of the PSF, even small artifacts 
can lead to spurious signals. We summarize the principal best practices 
extracted from the work of these authors:

\begin{enumerate}
\item Observe the source twice at each desired PA - once at the original 
position and then with the slit rotated 180${\degr}$. Instrumental 
artifacts should rotate with the instrument. If they are reproducible, 
then they should cancel when the signals are combined. 

\item Observe a standard star as a secondary check for instrumental artifacts

\item Acquire very high signal to noise ratio flats every time the grating is 
moved. If the grating in the spectrograph moves when the telescope slews, 
it may be necessary to retake flats with each observation. Small errors in 
the flat-fielding can shift the center of the PSF and result in spurious signals. 

\item The slit width should be narrow compared to the PSF. If the PSF 
does not fill the slit and the PSF is not perfectly symmetric, it can result 
in a spurious signal that will not be removed by combining with the SA 
signal acquired with the instrument rotated 180${\degr}$. 

\item Acquire spectra over a wide spectral range. It is crucial that the 
continuum be well sampled to ensure a baseline measurement is 
available for comparison to the region covered by the spectral line. 
This can prove to be challenging when observing strong \ion{H}{1} 
lines from Herbig Ae/Be stars with high resolution spectrographs 
as the the full width at zero intensity of the line can be comparable 
to the spectral grasp of the order. 

\end{enumerate}

 \begin{figure}
 \includegraphics[width=2.75in]{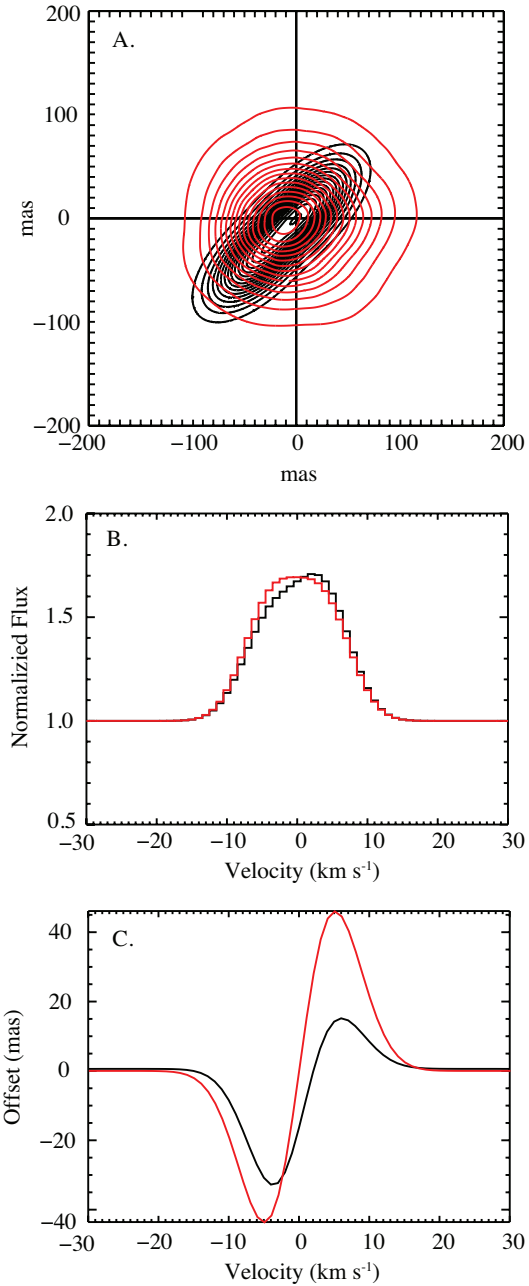}
 \caption{Illustration of the effect of an asymmetric PSF. In panel A, the PSF 
 of a star acquired with PHOENIX on Gemini South is plotted (red contours). 
 A synthetic, elongated profile is also plotted (black contours). An emission line was 
 synthesized assuming it arose from gas in a disk in a Keplerian circular orbit.  In panel B the 
 resultant line profiles that resulted from the PSF plotted in panel A are plotted. In panel C 
 the spectro-astrometric signal
 is similarly plotted. While the focus of the PSF has only a minor effect on the spectral line
 profile, the effect on the measurement of the spectro-astrometric signal is significant. Proper 
 focusing of the source is important for acquiring reliable signals. } 
 \label{fig:6}
 \end{figure}

In what follows, we highlight two additional issues that may complicate 
the interpretation of spectro-astrometric measurements. First we 
discuss the challenges of analyzing spectro-astrometric signals 
from resolved sources and second we explore the challenges of 
making SA measurements from spectra acquired with an IFS. 

When observing molecular emission from nearby disks, it is possible that the 
emission will be spatially resolved. For example, the molecular NIR emission 
observed from HD~100546 extends from $\sim$10$\pm2$~AU to beyond 
50~AU (van der Plas et al. 2009; Brittain et al. 2009; Carmona et al. 2011; 
Liskowsky et al. 2012). In addition there is a 
compact ring of solids that extends from 0.25~AU to less than 0.7~AU that 
dominates 
the continuum in the NIR (Benisty et al. 2010; Mulders et al. 2011, 2013; Pani{\'c}  et al. 
2014). 
At the distance to HD~100546 (97$\pm$4~AU; van Leeuwan 2007), the inner 
hole 
of the disk subtends 0$ \arcsec .28\pm0\arcsec.03$ along the semimajor axis. 
Adopting 
an inclination of 42$\degr$ (Ardila et al. 2007; Pineda et al. 2014) indicates 
that the 
inner hole subtends 0$\arcsec.19\pm0\arcsec.03$ along the minor axis of 
the disk. 

Spectroscopic observations of 
molecular lines from the disk using adaptive optics and $0\arcsec.2$ slit will 
result in the occultation of a significant fraction of the disk. Small changes in the 
positioning of the slit can  give rise to varying line shapes (Hein Bertelsen et al. 2014). 
Reconstruction of the line profile representing 
the underlying structure of the disk thus requires precise knowledge of the slit position. Even 
pointing uncertainties of order $0\arcsec$.05 can undermine the interpretation 
of the spectrum. In addition, since the continuum is unresolved, the PSF of the 
continuum and the PSF of the emission lines is not identical. This can give 
rise to artifacts as well since the PSF does not fill the slit.

In such cases it can be advantageous to work at lower spatial resolution and 
wider slit widths. 
Here we compare the effect of slight pointing offsets on the shape of the 
spectral lines to the effects 
described by Hein Bertelsen et al. (2014). We consider conditions with 
seeing of 0$\arcsec$.6 and 
a four pixel slit with a width of 0$\arcsec$.34 (the slit width of Phoenix on 
Gemini South; \citealt{2003SPIE.4834..353H,2000SPIE.4008..720H,1998SPIE.3354..810H}).  
If the slit is offset from the center of the PSF, an asymmetric PSF and spectro-astrometric signal does result 
(figure 5); however, to achieve 
a deviation of 20\% from the original measurement, the PSF of the continuum 
must be shifted 
as much as 
2 pixels  from the center of the slit. 
With care the star can be centered on the slit to within $\pm$0.5pixels
(about $\pm$0$\arcsec$.05 for Phoenix on Gemini South and CRIRES on the VLT). 
While such an offset can be significant when using adaptive optics (a PSF 
FWHM$\sim 0\arcsec.15$) and $0\arcsec.2$ slit (Hein Bertelsen et al. 2014)
it is much less significant when the seeing is $0\arcsec.6$. 

An additional concern arises from making measurements of 
spectro-astrometric signals from an uncorrected PSF. For example, an
asymmetric PSF can give rise to a spurious spectro-astrometric signal. We modeled 
the effect of a poorly focused beam convolving our disk model with an observed 
PSF and an asymmetric PSF (figure 6a). The line profile shows only a modest effect
due to the occultation of part of the disk (figure 6b). The spectro-astrometric signal, however, 
shows a significant asymmetry (figure 6c). 
Perhaps of greater concern is that rotating the instrument will not remove artifacts
arising from this effect if it is not steady (say due to variable seeing). 
Imaging the PSF throughout the observations is a helpful
way to identify problems with a variable PSF.

\section{Conclusions}
\label{sec:conclusions}
Spectroscopy has been used to elucidate many aspects of Herbig Ae/Be
stars since the previous conference on Herbig Ae/Be stars held in 1994; however, there is much exciting work
left to be done. Over the past two decades, we have moved from detecting and characterizing line emission arising 
from circumstellar material to using these features to elucidate the star-disk connection and identify signposts
of forming gas giant planets. The application of spectro-astrometric measurement to the study of disks is still in its 
infancy. While this technique shows remarkable promise for revealing sub-$milli$arcsecond spatial information 
from spectra, attention is needed to avoid several pitfalls that can complicate the interpretation of such data. The
detection of companions with orbits of order $\sim$10~AU by measuring the Doppler shift of emission arising
from the circumplanetary disk requires observations on decadal timescales. Archival data is thus a crucial resource
facilitating the long term monitoring of these systems.

%
\acknowledgments
S.D.B.  acknowledges support for this work from the National Science Foundation under grant number AST-0954811. Basic research in infrared astronomy at the Naval Research Laboratory is supported by 6.1 base funding.


\end{document}